**Research Article** Vol. 29, No. 10 / 10 May 2021 / *Optics Express* 14963

# Transverse mode instability and thermal effects in thulium-doped fiber amplifiers under high thermal loads

CHRISTIAN GAIDA,[1,2,*] MARTIN GEBHARDT,[1,3] TOBIAS HEUERMANN,[1,3] ZIYAO WANG,[1] CESAR JAUREGUI,[1] AND JENS LIMPERT[1,2,3,4]

[1]*Institute of Applied Physics, Abbe Center of Photonics, Friedrich-Schiller-Universität Jena, Albert-Einstein-Str. 15, 07745 Jena, Germany*
[2]*Active Fiber Systems GmbH, Ernst-Ruska Ring 17, 07745 Jena, Germany*
[3]*Helmholtz-Institute Jena, Fröbelstieg 3, 07743 Jena, Germany*
[4]*Fraunhofer Institute for Applied Optics and Precision Engineering, Albert-Einstein-Str. 7, 07745 Jena, Germany*
[*]*gaida@afs-jena.de*

**Abstract:** We experimentally analyze the average-power-scaling capabilities of ultrafast, thulium-doped fiber amplifiers. It has been theoretically predicted that thulium-doped fiber laser systems, with an emission wavelength around 2 µm, should be able to withstand much higher heat-loads than their Yb-doped counterparts before the onset of transverse mode instability (TMI) is observed. In this work we experimentally verify this theoretical prediction by operating thulium doped fibers at very high heat-load. In separate experiments we analyze the performance of two different large-core, thulium-doped fiber amplifiers. The first experiment aims at operating a short, very-large core, thulium-doped fiber amplifier at extreme heat-load levels of more than 300 W/m. Even at this extreme heat-load level, the onset of TMI is not observed. The second experiment maximizes the extractable average-output power from a large-core, thulium-doped, fiber amplifier. We have achieved a pump-limited average output power of 1.15 kW without the onset of TMI. However, during a longer period of operation at this power level the amplifier performance steadily degraded and TMI could be observed for average powers in excess of 847 W thereafter. This is the first time, to the best of our knowledge, that TMI has been reported in a thulium-doped fiber amplifier.

## 1. Introduction

Fiber laser systems are known for their excellent thermal-handling capabilities and output beam quality, even at very high average-output powers. Such a reputation has been mostly attained thanks to the development of high-power ytterbium-doped fiber lasers (YDFL), which are well-known for their low quantum-defect heating and their very high emission efficiency at a wavelength of around 1 µm [1,2]. This, together with the availability of cost-effective, high-brightness diode-laser technology around 976 nm for cladding pumping, was key for demonstrating kilowatts of average-power in CW operation with single Yb-doped fiber emitters [3–6]. However, in spite of the early successes, the average power scaling of YDFLs (single emitters) has stagnated in the last decade and, in particular, ultrafast CPA schemes have only recently crossed the kilowatt level [7–9]. This is because an unexpected average-power-limiting effect was discovered: transverse-mode instability (TMI) [10]. Due to TMI, a formerly stable Gaussian-like beam emitted by a high-power fiber laser system becomes transversally unstable once that a certain average power-threshold is reached [11]. Further investigations revealed that this instability is caused by a dynamic energy transfer between fiber modes [12] and that the onset of this effect is caused by a thermally-induced index grating. Over the last decade TMI has been intensively investigated experimentally and theoretically for YDFLs [13–16]. TMI is typically observed for average heat-loads < 35 W/m [17] (please note that this is a soft upper limit for many conventional systems, but it can be shifted depending on the experimental conditions, e.g. gain saturation [18]). Even though this is, by no means, a hard condition, it gives a notion of the magnitude of the heat-load required to reach the TMI threshold in YDFLs and the difficulty of obtaining several hundred Watts of average power from short-length, rod-type fiber amplifiers, which are desired for high-energy fiber CPAs.

Interestingly, there is theoretical [19] and experimental [20] evidence suggesting that thulium-doped fiber lasers (TDFL) can withstand much higher heat-loads than their Yb-doped counterparts without reaching the TMI threshold. However, this potential advantage has not yet resulted in an average-power scaling beyond the limits already observed with YDFLs. The reason for this is that power-scaling TDFLs faces severe challenges associated with the complex energy-level structure and the lower efficiency of thulium-doped fused silica, the lack of high-power pump diodes and optical components as well as the atmospheric water-vapor absorption lines that overlap with the gain-bandwidth of thulium [21]. In particular the lower efficiency leads to a significantly enhanced heat-load upon amplification in comparison to YDFLs, which is the primary obstacle for average-power scaling. Thus, the average power demonstrated with TDFLs has been limited to the 1 kW-level in continuous wave operation [20,22,23] and to the 100 W-level in ultrafast systems [24,25]. Only recently we demonstrated an ultrafast TDFL with an average output power of more than 1 kW with near diffraction-limited beam quality [26]. All these experiments operated below the theoretical efficiency maximum. Therefore, there is room for an optimization of the amplifier efficiency, which will lead to higher average power. In particular for TDFLs this means that the beneficial contribution of cross-relaxation is desired [27], which requires a careful optimization of the thulium-doped fiber and of the experimental conditions. So far, most studies have focused on optimizing the doping concentration of thulium-doped fiber amplifiers (TDFAs) [28,29], which has led to the demonstration of slope efficiencies around 50–60% [20,22,23]. On the other hand, even at comparable doping concentrations, significantly lower slope efficiencies (of around 10–40%) have been reported for short-length, very-large-mode- area (VLMA) TDFAs, which are required to reduce the impact of nonlinearity in pulsed TDFLs [30–32]. This highlights the fact that not only the ion concentration, but also the operating conditions play an important role in determining the cross-relaxation rate.

In this contribution, we investigate the onset of thermal effects in TDFLs under very high heat-load. In a first experiment we study the performance of a short-length VLMA TDFA at various heat-load levels, demonstrating a record average heat-load of 300 W/m without the onset of TMI. In a second experiment we present a large-mode area TDFA that reaches more than 60% optical-to-optical efficiency and an average-output power of 1.15 kW. During a longer period of operation at this power level, a progressive degradation of the amplifier performance occurred, and, to the best of our knowledge, TMI was observed in a TDFA for the first time.

## 2. High average-power operation of a rod-type fiber amplifier

This section describes a detailed experimental investigation on the average-power scaling of a rod-type, thulium-doped, large-pitch-fiber amplifier (Tm:LPF45). In the past, this fiber type has been used in CPA systems emitting ultrafast pulses with record-high peak-powers of several GWs [33]. However, the achievable average-power in the reported systems was hampered by thermal-blooming due to atmospheric water-vapor absorption [21] and the limited availability of high-brightness diode lasers for optical pumping at 790 nm wavelength. By circumventing these obstacles in this investigation, the average-power-scaling capabilities of Tm-doped rod-type fiber amplifiers can be finally assessed.

*2.1. Experimental setup*

The high average-power experiment was carried out in a standard master-oscillator, power-amplifier configuration, which is schematically illustrated in Fig. 1. The ultrafast seed source is the TDFL that has been presented and described in detail in Ref. [26]. It provides linearly polarized, stretched pulses with a full-width at half-maximum duration of 600 ps at a pulse repetition frequency of 80 MHz covering a spectral range between 1930–1990 nm. Several pre-amplifiers boost the average power to about 100 W. This average-power level causes thermal lensing effects in the currently commercially available Faraday isolators and, therefore, an alternative isolation scheme has to be implemented. A combination of a thin-film polarizer (TFP) and a quarter-wave plate (QWP) are used to protect the high-power seed system from any direct back reflections.

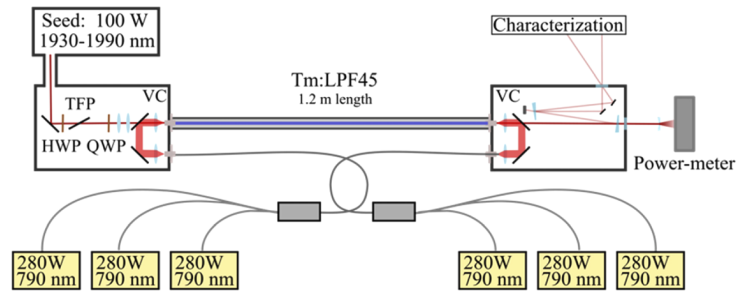

**Fig. 1.** Schematic illustration of the experimental setup for the amplification of stretched ultrafast pulses in the 2 µm wavelength region to high average-power in a thulium-doped large-pitch fiber (Tm:LPF45). TFP – thin-film polarizer; HWP – half-wave plate; QWP – quarter-wave plate; ; VC – vacuum chamber.

The main amplifier is a 1.2 m long Tm:LPF45 which has a pitch (hole-to-hole distance) of 45 µm and an air-hole-diameter-to-pitch ratio of 0.18. The fundamental mode (FM) of this fiber has a cold mode-field diameter (MFD) of 87 µm. At moderate heat-load levels, the Tm:LPF45 emits near diffraction-limited light with a typical M-squared value of <1.3. An air-cladding structure with a diameter of 270 µm allows for the guidance of low-brightness 790 nm pump light provided by commercially available laser-diodes. The small-signal cladding pump-absorption of this fiber is 12.3 dB/m at 790 nm wavelength. In this system state-of-the-art laser-diode modules at 790 nm wavelength provide a power of approximately 280–300 W each, which are delivered by multi-mode fibers with a core diameter of 200 µm and an NA of 0.22. In order to launch an even higher pump-power to the fiber amplifier, 3 × 1 pump combiners are used. As a result, up to 770 W of combined pump power can be delivered by a single multi-mode fiber with a core diameter of 400 µm and an NA of 0.22. Thus, pumping the Tm:LPF45 from both sides allows for a maximum pump power of 1540 W. The Tm:LPF45 is equipped with anti-reflection coated, fused-silica end-caps and is directly water cooled. All free-space sections in the setup are enclosed in vacuum chambers which are first evacuated and then filled with 1 bar of dry helium gas, as this efficiently mitigates atmospheric-water absorption effects [21]. The amplified signal is sampled by two fused-silica wedges, which allow for a low average-power beam characterization outside of the helium-chamber in normal atmosphere. The transmission through the first anti-reflection coated wedge is directly measured by a thermal power-meter.

*2.2. Absence of transverse-mode instability at extreme heat-load levels*

The measured power-slope of the TmLPF:45 is depicted in Fig. 2(a). The average slope efficiency amounts to 42% when a pump-coupling efficiency of 90% is assumed. The slope is nearly linear up to an average-output power of 500W and degrades slightly at higher powers. The relatively low efficiency indicates nearly no net contribution of cross-relaxation in the amplification process, which can be partly explained by the large mode-field area of the fiber that leads to a low signal saturation level. Additionally, the high pump absorption and short length of the fiber promotes a high inversion level that weakens the positive contribution of cross-relaxation and strengthens the opposite effect of energy-transfer up-conversion, which can in principle reduce the efficiency even below the stokes limit. The inversion level along the fiber is also influenced by the temperature dependence of the emission and absorption cross-sections. At high temperatures, i.e. high heat-loads at very high average powers, this can additionally reduce the efficiency [34,35]. Moreover, higher-order modes (HOMs) can lead to a reduced saturation level along the fiber amplifier and can further diminish the efficiency, as is observed in this amplifier at average output-powers >500 W.

The amplification process has been simulated assuming a seed-coupling efficiency to the fundamental mode of 70%. The simulation tool is based on the steady-state solution of the rate equations of the first four levels in thulium-doped fused silica [36,37] and includes thermal effects such as the mode-field-area shrinking and the temperature dependence of the emission and absorption cross-sections in thulium-doped fused silica. The simulation matches the experimentally measured power-slope very well for average-output powers of up to about 500 W. Beyond that point the model slightly overestimates the amplifier performance. This discrepancy likely originates from the existence of HOMs in the fiber core at high heat-load levels, which are not considered by the model. The HOMs can actually reduce the effective overlap of the signal field with the doped-core region and, therewith, the efficiency of the amplifier. The simulation predicts a transmitted pump power of at least 140 W at each side of the fiber. Thus, a further increase of the pump power (e.g. by utilizing $4 \times 1$ pump combiners) poses a high risk for damaging the pump combiners and laser diodes.

The beam stability, i.e. the potential onset of TMI, is monitored following the procedure described in Ref. [38]. Hereby the output beam diameter is expanded to about 5 mm and is directed to an InGaAs photodiode that has a detector diameter of 0.5 mm. Since the beam is much larger than the detector, any fluctuation in the spatial beam profile will ultimately lead to an intensity variation on the detection area of the photodiode. The suppression of the pulse repetition frequency is achieved with a 5 MHz electronic low-pass filter. The photodiode signal is recorded for various average output powers over a duration of 2 s with a 12-bit oscilloscope and a total of 25 mega-samples. The power-spectral density (PSD) of the detected fluctuations is obtained by performing a Fourier-transformation of each temporal trace after normalization to the DC-value and the sampling rate. In order to evaluate the evolution of the root-mean-square of the power fluctuations with increasing optical average power, the PSDs are integrated over a frequency band ranging from 10 Hz to 1 MHz for all power levels. According to the definition provided in Ref. [38], the TMI threshold is reached once the derivative of the average-power-dependent, RMS of beam fluctuations exceeds 0.1 ‰/W. The measured RMS of beam fluctuations with respect to the average-output power is depicted in Fig. 2(b)) and increases slightly with the rising average-output power. It is likely, but cannot be confirmed with certainty at this point, that the slight increase of the RMS beam fluctuations represents the beginning stages of TMI. The measured evolution of the RMS beam fluctuations has been fitted (red line in Fig. 2(b))) in order to determine its derivative (blue line in Fig. 2(b))), which remains well below 0.1 ‰/W even at the highest output power. Although it is likely that a further increase of the pump power would reveal the TMI threshold of this fiber, we refrained from increasing the pump power any further considering the substantial risk of damaging the pump diodes due to a significant transmission of the pump power at the highest power levels (as discussed above).

Although the TMI threshold of the TmLPF45 has not been reached even at the highest launched pump-power, a static deviation from the fundamental-mode profile is nevertheless observed.

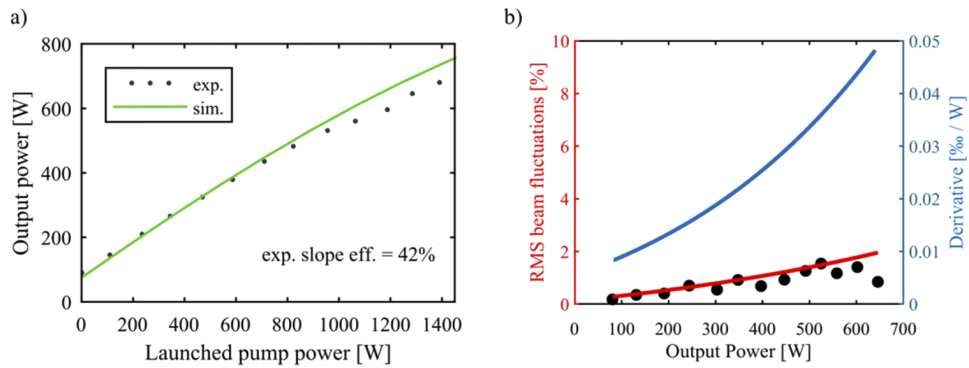

**Fig. 2.** a) Measured (exp.) and simulated (sim.) power slope of the Tm:LPF45. b) RMS beam fluctuations that indicate the beam stability vs. average-output power. The red line represents the fit of the RMS beam fluctuations and the blue line its calculated derivative, which is found to be smaller than 0.1 ‰/W at all measured power levels.

Fig. 3(a) shows the emitted beam profiles at the fiber-output facet for several average-powers. These images reveal a significant heat-load-induced shrinking of the mode-field area (MFA) with increasing average-output power. This remains the dominant effect up to an average-output power of approximately 400 W where the onset of significant static distortions of the output-beam profile can be observed. A further analysis reveals that the onset of the beam distortion is accompanied by a spectral modulation, as shown in Fig. 3(b)). This modulation reveals the interference of at least two fiber modes. Even though the lack of systematic investigations, i.e. an analysis of the mode content at different average-output powers, makes the claim rather speculative at this point, note that such a static beam degradation might be the result of modal coupling induced by the strongly longitudinally asymmetric thermally-induced index grating in this fiber, as has been predicted elsewhere [39,40].

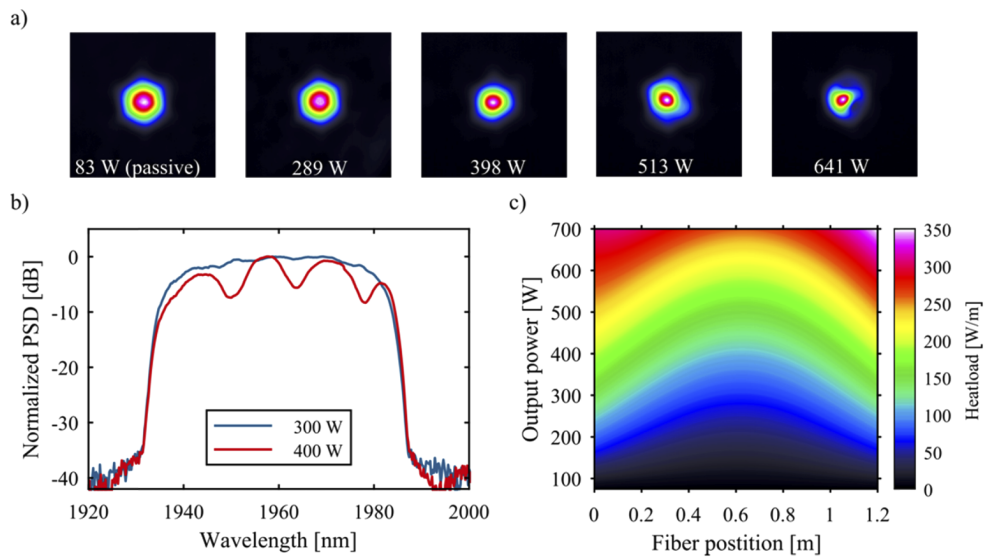

**Fig. 3.** a) Beam images at the fiber output for different average-output powers. Figure 3(b)) Spectral modulations are evident at power-levels ≥ 400 W. Figure 3(c)) Simulated heat-load in the fiber core with respect to the average-output power and the longitudinal fiber position.

Using the simulation of the amplification process, the heat-load evolution along the fiber can be calculated at any given pump power. At the highest average-power, a peak heat-load of 350 W/m and an average heat-load of 300 W/m have been determined, which are about an order of magnitude higher than the typical heat-load-threshold for TMI in YDFLs. Please note, that these calculated values are based on the simulation of the amplification process as is shown in Fig. 2(a) and considering the small deviation from the measurement results, the real values might deviate slightly.

## 3.　Observation of TMI in a thulium-doped fiber amplifier

In a recently reported experiment, an ultrafast TDFL reached an average output power of more than 1 kW with near diffraction-limited beam quality [26]. The main amplifier in this experiment was built using a polarization-maintaining, thulium-doped, photonic-crystal fiber [41] (Tm:PCF50/250) with a total length of 6 m. The slope efficiency of >60% in this experiment reveals the beneficial contribution of cross-relaxation processes when pumping the amplifier at 790 nm wavelength. The reasons for the improved cross-relaxation rate of the Tm:PCF50/250 as compared to the rod-type Tm:LPF45 (described in the previous section) lie mainly in the stronger signal saturation and increased fiber length. This way, it is expected that the impact of thermal effects is shifted to significantly higher average powers in the Tm:PCF50/250 than in the Tm:LPF45.

### 3.1.　Experimental setup

A sketch of the experimental setup for the amplification experiment in a Tm:PCF50/250 is illustrated in Fig. 4. The setup is, in principle similar, to the amplification experiment in section 2 with the exception that the seed average-power is reduced to 10 W in order to allow for the use of a Faraday-isolator. The improved isolation is required in order to prevent parasitic lasing that can arise due to the high gain of the long fiber-amplifier.

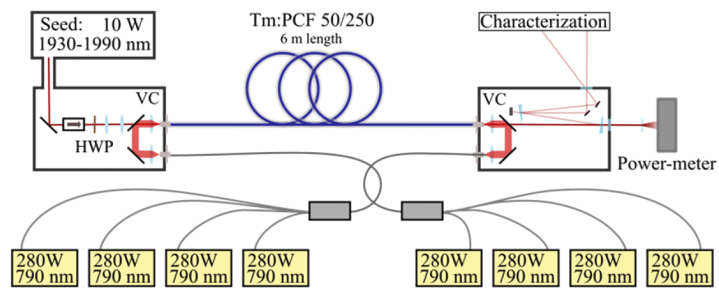

**Fig. 4.** Schematic illustration of the experimental setup for the high-power amplification of ultrafast pulses in the 2 µm wavelength region using a thulium-doped photonic-crystal fiber (Tm:PCF50/250). HWP – half-wave plate; VC – vacuum chamber.

The mode-field diameter (MFD) of the fundamental mode of the polarization-maintaining Tm:PCF50/250 is 35 µm with a typical M-squared value of <1.1 [26,41]. The air-cladding structure of this fiber has a diameter of 250 µm. The pump is free-space coupled to the cladding with a specially designed pump-coupling arrangement. Hereby, eight pump modules, each delivering 280 W of CW power at 790 nm, are efficiently merged with 4 × 1 pump combiners. This way a maximum pump power of 2090 W is available for bi-directional pumping. The pump transmission at high-power operation can be minimized by choosing a fiber length of 6 m (since the small-signal cladding-pump absorption is 6.2 dB/m at 790 nm wavelength). Moreover, the Tm:PCF50/250 is coiled to a diameter of approximately 25–30 cm in order to ensure robust single-mode operation, a procedure that is known to increase the TMI threshold [42]. Similar to the setup described in section 2, the Tm:PCF50/250 is equipped with anti-reflection coated, fused-silica end-caps and is directly water cooled.

*3.2. Observation of transverse-mode instability after amplifier degradation*

The first measurement run is done with a pristine fiber, i.e. the launched pump-power in this fiber prior to this experiment never exceeded 50 W. All results in Fig. 5 and Fig. 6 that correspond to measurements with the pristine fiber are shown in blue. The pump-power is increased in several steps to the maximum available level (2090 W), which corresponds to 1881 W launched-pump power when assuming 90% pump coupling efficiency to the main amplifier. The pristine-amplifier slope with respect to the launched pump-power is shown in Fig. 5(a). The pristine amplifier delivers up to 1.15 kW of average power at the maximum launched pump-power level and has a slope efficiency of >60%. Even at the highest power-level, the beam quality is nearly diffraction limited [26]. The simulation result of the amplification process is also shown in Fig. 5(a) (orange line) and proves very good agreement with the measurement. Similar to the simulation procedure in section 2 of this paper we retrieve the heat-load evolution along the pristine fiber amplifier, which is depicted in Fig. 5(b), and find a peak heat-load of >200 W/m at the fiber output. The average heat-load amounts to 95 W/m at the highest average-output power.

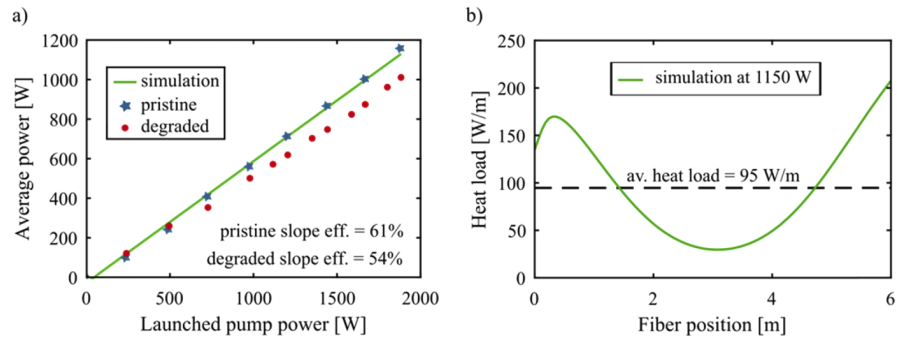

**Fig. 5.** a) Measured average-output power with respect to the launched pump power for the pristine fiber (blue stars) and the degraded fiber (red circles). The slope efficiency degrades from 61% to 54% after 60 min of continuous operation at >800 W average-output power. The simulation (green line) is in very good agreement with the measured slope of the pristine fiber. b) Simulated heat load evolution along the pristine fiber at the maximum average-output power. The average heat load reaches 95 W/m.

In order to characterize the amplifier with respect to TMI, the beam stability is characterized following the same procedure as in section 2 of this paper [38]. As shown in Fig. 6(a) the measured PSD of beam fluctuations of the pristine fiber at 1.15 kW average-output power verify the high beam stability of the amplifier. Fig. 6(b) shows the RMS of beam fluctuations being as low as 0.32% even at the highest power-level and there is no evidence of TMI in the pristine fiber. This result is qualitatively equivalent to the findings of our recently published work [26].

After about 20 minutes of continuous operation at the maximum average-output power of 1.15 kW, the onset of TMI is observed. The red line in Fig. 6(a) exemplarily illustrates the PSD of the beam fluctuations when operating the degraded fiber at 900 W average-output power. In contrast to the blue line shown in the same figure, which is the measured PSD of the pristine fiber at 1.15 kW average-output power, strong beam fluctuations are found in the frequency range 100 Hz- 3 kHz. The output beam is additionally analyzed with a camera that captures 100 frames per second (see Visualization 1). The video reproduces the frames at a reduced rate of 5 fps in order to visualize the dynamic changes of the beam profile at the output facet of the fiber, which is imaged onto the camera chip and showing clear evidence on the involvement of the LP11 mode as expected with the onset of TMI. The red line in Fig. 6(b) shows the RMS of beam fluctuations with respect to the average-output power of the degraded fiber. The dashed vertical line corresponds to the power level at the TMI threshold, where the derivative of RMS of beam fluctuations reaches 0.1 ‰/W. Within 60 minutes of continuous operation at an average-power >800 W, the TMI-threshold degrades to a final value of 847 W. Interestingly, the degradation of the TMI-threshold is accompanied by a degradation of the amplification efficiency. The red line in Fig. 5(a) represents the measured power slope of the degraded fiber and reveals a reduction of the efficiency to 54%.

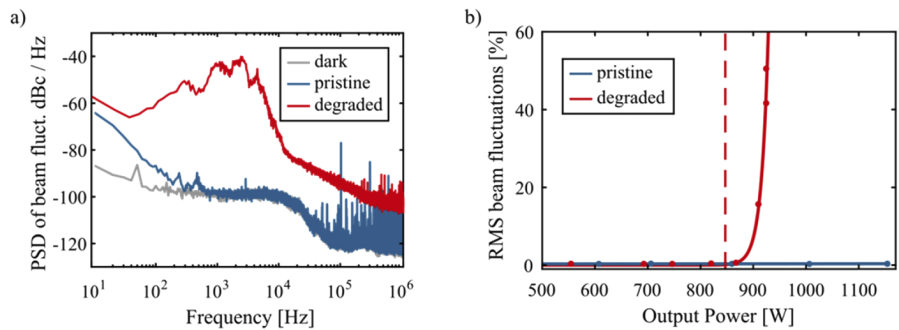

**Fig. 6.** a) PSD of beam-intensity fluctuations for the pristine fiber (blue line) and for the degraded fiber (red line). Significant noise is observed in the frequency range 100 Hz- 3 kHz for the degraded fiber. B) RMS of beam fluctuations with respect to the average-output power of the pristine fiber (blue line) and of the degraded fiber (red line).

The reduced TMI-power threshold suggests either the formation of permanent waveguide changes and/or an increased heat-load level at a given average power. Clearly, the current experimental investigation results are insufficient to identify the underlying reason for the observed degradation. Therefore, the aim of the following two paragraphs is to elaborate on possible mechanisms that could be the cause for the observed degradation to establish the basis for further investigations in the future.

As mentioned above, the energy coupling to the HOMs can be the result of permanent waveguide changes. Such permanent refractive-index modifications have been observed in untreated fused-silica fibers after illumination with intense UV radiation [43]. The magnitude of the observed modification is on the order of $\Delta n = 10^{-5} \ldots 10^{-4}$, which is certainly comparable to the typical thermally induced refractive-index changes [44]. UV-radiation from thulium-doped fused silica under 790 nm pumping is possible but requires a considerable population of higher lying energy levels, from where the UV-emission (290–350 nm) originates [27]. In order to verify or disprove the involvement of this mechanism, an accurate evaluation of the intensity from the thulium-ions in the active core of the Tm:PCF50/250 and an a comparison to known energy-transfer mechanisms is necessary.

Photodarkening has been identified as an additional loss mechanism and heat source in YDFAs, which can significantly reduce the average-power threshold of TMI [45]. It is generally accepted that the additional losses are related to the formation of color centers (CC) [46]. Photodarkening has also been reported in thulium-doped aluminosilicate fibers [47–49] and the degradation is commonly believed to originate from upconversion-induced, short-wavelength emission of the thulium ion. Another process that might be correlated to photodarkening in TDFA is a reduction of the cross-relaxation rate, which significantly contributes to the overall efficiency and heat-load upon amplification. The formation of CCs might reduce the cross-relaxation rate due to the deactivation of thulium-ions similar to the proposed degradation mechanism in YDFA [50]. A detailed study of photodarkening in high-power TDFAs will be required in the future to shed light on the degradation mechanisms observed in our experiment.

## 4. Conclusion and outlook

In this work we have investigated thermal effects in thulium-doped VLMA fiber amplifiers under extreme heat loads with two separate experiments. As a general result, we can conclude that TDFLs have a much higher resilience against thermally-induced waveguide changes than YDFLs.

In the first experiment we study the performance of a short-length, rod-type TDFA and deduce from the measurements a record average heat load of 300 W/m without the onset of TMI. Although we find evidence of HOMs at average-power levels >400 W, the associated output-beam deformations are static. This behavior could be pre-compensated with an adapted fiber design [51,52], allowing for fundamental mode operation even at extreme heat-load levels, which can prospectively pave the way to kW-level average powers from rod-type TDFAs. This prospect is of particular interest for ultrafast TDFLs that aim at the simultaneous emission of high peak- and average-power in combination with a close-to-diffraction-limited beam quality – which represents desired laser parameters of many visionary applications.

In a second experiment we demonstrate TMI-free 1150 W of average output power from a flexible TDFA with a core diameter of 50 µm. Considering the average heat-load of 95 W/m that has been deduced from the experimental measurements, a significantly higher heat-load robustness as compared to YDFLs is experimentally verified. In spite of this promising result, the long-term operation of the current system at high average power level reveals a degradation of the amplifier performance. This deterioration includes a decreased slope efficiency and a reduced threshold for TMI. These effects were reported in TDFLs for the first time, to the best of our knowledge. Further experimental and theoretical investigations of the observed effects are necessary to optimize TDFLs and to fully exploit their power-scaling potential. We have outlined possible directions for further investigations to reveal the underlying mechanisms of the observed degradation.

In summary, the realization of the presented high-average power TDFLs unveils the power scaling potential of this technology. It has been demonstrated that it holds the great promise to uniquely combine the emission of very high average and peak power at around 2 µm wavelength from a single ultrafast fiber amplifier. These laser parameters are of outmost interest for many demanding ultrafast applications, e.g. high-harmonic generation [53], particle acceleration [54], mid-IR generation [55], as well as life sciences and material processing.

**Funding.** Deutsche Forschungsgemeinschaft (416342637, 416342891); Fraunhofer Cluster of Excellence Advanced Photon Sources (CAPS); European Research Council (835306).

**Acknowledgments.** We acknowledge Dr. Christoph Stihler for his advice and help with the experimental characterization of the TMI-threshold. This work was supported by the European Research Council (ERC) under the European Union's Horizon 2020 research and innovation program (grant 835306, SALT) and the Fraunhofer Cluster of Excellence Advanced Photon Sources (CAPS). C.J. acknowledges funding by the Deutsche Forschungsgemeinschaft (DFG, German Research Foundation) - 416342637; 416342891.

**Disclosures.** The authors declare that there are no conflicts of interest related to this article.